\documentstyle[preprint,aps]{revtex}
\begin{document}
\draft
\title{Charge quantization in a chiral bilepton gauge model} 
\author{ C.A de S. Pires and O.P. Ravinez}
\address{\tightenlines{Instituto de  F\'{\i}sica Te\'{o}rica, Universidade 
Estadual
Paulista, Rua Pamplona 145, 01405-900 S\~{a}o Paulo, S\~{a}o Paulo, 
Brazil.}}
\maketitle
\begin{abstract}

 In the context of the standard model the quantization of the electric charge  
occurs only family by family. When we  consider the three families together  
with massless neutrinos the electric charge is not quantized any more. 
Here we show  that a chiral 
bilepton gauge model based on the gauge group $SU(3)_C\otimes 
SU(3)_L\otimes U(1)_N$  explains  the quantization of the electric 
charge when we take into account the three families of fermions. This result
does not depend on the neutrino masses. Charge quantization occurs either the
neutrinos are massless or Dirac or Majorana massive fields. 
\end{abstract}
\section{INTRODUCTION}

The reason why the electric charges of the fundamental particles 
appear only in discrete units is still an open question. In the 
course of the  years some proposals have appeared intending to 
explain 
it. 
The first proposal was given by Dirac\cite{1} through the postulate of the 
magnetic monopoles. The second proposal comes from the grand 
unification theories (GUT'S) through its group structure\cite{2}. But the GUT'S 
are ruled out and magnetic monopoles have not still been detected.

Recently the quantization of the electric charge (QECH)  has been analyzed 
within the gauge models that contain the $U(1)$ factor in its gauge 
group\cite{3,4,5}. The approach given here relates the $U(1)$ charges of the  
fermions and Higgs of the 
model through  classical and quantum constraints in such a manner 
that it
leads to the QECH. The classical 
constraints imply the Lagrangian of the model be invariant by the gauge 
group; the quantum ones imply  the model be free from anomalies\cite{3,4,6}.

Analyzing from this point of view the QECH in the standard 
model (SM), whose gauge group is $G_{SM}=SU(3)_C \otimes SU(2)_L \otimes 
U(1)_Y$, 
several authors showed 
that the SM with one family contains the QECH\cite{3,4,5,7,8}. Nevertheless, 
when we increase the number of families to three the effect of dequantization 
occurs\cite{3,8,9}. To understand this we need to see that in the SM 
with three families  an independent anomaly free global $U(1)_{Y_1}$ symmetry 
arises\cite{3,4}. By independent we mean that the 
$U(1)_{Y_1}$ is independent  of the gauge symmetry of the SM. By anomaly free 
we mean that the $U(1)_{Y_1} G_{SM}^2$ and $U(1)^3_{Y_1}$
anomalies are canceled\cite{3}. This kind of symmetry is also called 
hidden symmetry. It creates an arbitrariness in the definition  
of the electric charge operator since $Y$ and $Y +\alpha Y_1 $ are equally good 
choices for the gauge $U(1)$ in the gauge group of the standard 
model\cite{3,4}.

We show using this approach  that in a chiral bilepton gauge model, 
proposed some years ago by  Pisano,  Pleitez and  Frampton \cite{10}, based
 
on the symmetry gauge group   $ G_{331} = SU(3)_C\otimes SU(3)_L\otimes U(1)_N 
$, there is the QECH through classical and quantum constraints independently if 
neutrinos are massless or not. This model embeds the SM; it adds new physics 
with no hidden symmetries, the leptons come in the $SU(3)_L$ representation  and 
it
has three nontrivial anomaly cancellations\cite{11}. 

This paper is organized as follows. In Sec.II we analyze the QECH in the SM 
with one and three families  showing the hidden symmetries that lead to the 
effect of dequantization. In Sec.III we extend the analysis to the chiral 
bilepton gauge model and we  obtain the quantization of the electric charge 
through classical and quantum constraints and in Sec.IV we summarize our 
conclusions.

\section{The quantization in the standard model}

\subsection{The standard model with one Family}

The electric charge operator in the SM can be defined in a general form 
as:
\begin{equation}
Q=T_3 + b Y.
\label{1}
\end{equation}
Where $b$ is an unknow parameter. For having nonvanishing fermion masses we 
must introduce a Higgs doublet $ \phi \sim (\mbox{{\bf 1,2}},Y_\phi )$ 
that 
acquires a vacuum expectation value:
\begin{eqnarray}
\langle \phi \rangle_0 \sim
\left (
\begin{array}{c}
0 \\
v
\end{array}
\right ) . 
\label{2} 
\end{eqnarray} 

Since we want the operator $Q$ unbroken, $Q\langle \phi \rangle_0$ 
must be 
zero. With this condition we find $b=\frac{1}{2} Y_\phi$. So the 
electric 
charge operator takes the form:
\begin{equation}
Q=T_3 +\frac{Y}{2Y_\phi}.
\label{3}
\end{equation} 
Then, the problem of the quantization of the electric charge turns into 
writing all hypercharges as functions of the Higgs hypercharge $Y_\phi$. 
To 
achieve this we use the classical and quantum constraints. The 
only sector in the Lagrangian able to give information about the 
hypercharges through classical constraints is the Yukawa one, while the 
information 
about the hypercharges through quantum constraints comes from  the three 
nontrivial 
anomaly cancellations $[U(1)_Y]^3 $, $[SU(2)_L]^2 U(1)_Y $  and 
$[SU(3)_C]^2 U(1)_Y $.
 
In the SM the quarks and leptons come in the following representations:
\begin{eqnarray}
&&L_L=
\left (
\begin{array}{c}
\nu_e \\
e
\end{array}
\right )_L \,\, \sim (\mbox{{\bf 1,2}},Y_l),\,\,\,e_R \sim (\mbox{{\bf 
1, 
1}},Y_e),\nonumber \\
&&Q_L=
\left (
\begin{array}{c}
u \\
d
\end{array}
\right )_L \,\, \sim (\mbox{{\bf 3}}, \mbox{{\bf 2}},Y_q),\,\,\, u_R 
\sim 
(\mbox{{\bf 3,1}},Y_u),\,\,\, d_R \sim (\mbox{{\bf 3,1}},Y_d) ;
\label{4} 
\end{eqnarray}
with the Yukawa interaction being:
\begin{equation}
-{\cal L}^Y = g^l_1 \bar L_L \phi e_R + 
g^d_3 \bar Q_L \phi d_R + g^u_2 \bar Q_L \tilde \phi u_R + H.c..
\label{5}
\end{equation}
This Lagrangian, for being $U(1)_Y $ gauge invariant, implies that:
\begin{equation}
Y_e = Y_l -Y_\phi,\,\,\, Y_u = Y_q + Y_\phi,\,\,\, Y_d = Y_q - Y_\phi.
\label{6}
\end{equation}

After this only two nontrivial anomaly constraints remain: 
\begin{eqnarray}
&&[SU(2)_L]^2 U(1)_Y \Longrightarrow Y_q=-\frac{1}{3}Y_l,\nonumber \\
&&[U(1)_Y]^3_Y \Longrightarrow Y_l=-Y_\phi.
\label{7}
\end{eqnarray}

Eqs.(\ref{6}) and Eqs.(\ref{7}) leave all fermion hypercharges as 
functions of 
the Higgs one:
\begin{equation}
Y_l=-Y_\phi,\,\,\,Y_e=-2Y_\phi,\,\,\,Y_q=\frac{1}{3}Y_\phi,\,\,\,
Y_u=\frac{4}{3}Y_\phi,\,\,\,Y_d=-\frac{2}{3}Y_\phi.
\label{8}
\end{equation}
Substituting the above results into Eq.(\ref{3}), we obtain the  quantization of 
the 
electric charge with the correct electric charges for leptons and quarks:
\begin{equation}
 Q_\nu =0,\,\,\, Q_e=-1,\,\,\, Q_u=\frac{2}{3}\,\, \mbox{and}\,\, 
Q_d=-\frac{1}{3}.
\label{9}
\end{equation} 

Next if we admit a right-handed neutrino with Dirac mass term and 
we attribute the 
hypercharge $Y_\nu $ to it we find from the Yukawa term $\bar L_L \tilde 
\phi 
\nu_R$ that $Y_\nu = Y_l + Y_\phi$. But now we are only with  
one nontrivial anomaly constraint $[SU(2)_L]^2U(1)_Y$ . So we have three free 
parameters $Y_l$ , $Y_q$ and $Y_\phi$ from classical constraints and only one 
equation 
from quantum constraints. This prevents us 
from leaving all the hypercharges as functions of only the Higgs one. 
This is 
the dequantization effect\cite{3,4}. The explanation is that the SM  with 
Dirac 
massive 
neutrinos presents, besides the barionic (B) and leptonic(L) global 
symmetries, the 
$B-L$ global symmetry. The $B$  and  $L$ symmetries are not free from 
anomalies. So they 
are not hidden symmetries. Nevertheless, the $B-L$ symmetry is free from 
anomaly. 
Then it 
is a hidden symmetry which superposes to the 
hypercharge one and which obstructs us to know if the $U(1)_Y $ factor in the 
electric
charge operator is due to $Y$ or to the superposition $ Y + 
\alpha(B-L)$\cite{3,4}.  Now if we, instead of Dirac neutrinos, have a 
Majorana 
one with the mass-term $\nu^T_R C^{-1} \nu_R$ that breaks the $B-L$ global 
symmetry, we restore the quantization condition\cite{5}.

\subsection{ The standard model with three families}

With three families the representation content is:
\begin{eqnarray}
&&L_{aL}=
\left (
\begin{array}{c}
\nu_a \\
e_a
\end{array}
\right )_L  \sim (\mbox{{\bf 1,2}},Y_{l_a}),\,\,\,e_{aR} \sim 
(\mbox{{\bf 1, 
1}},Y_{e_a}),\nonumber \\
&&Q_{aL}=
\left (
\begin{array}{c}
u_a \\
d_a
\end{array}
\right )_L  \sim (\mbox{{\bf 3,2}},Y_{q_a}),\,\,\, u_{aR} \sim 
(\mbox{{\bf 
3,1}},Y_{u_a}),\,\,\, d_{aR} \sim (\mbox{{\bf 3,1}},Y_{d_a}) ;
\label{10} 
\end{eqnarray}
with the Yukawa interaction being:
\begin{equation}
-{\cal L}^Y = g^l_{aa}\bar L_{aL} \phi e_{aR} + g^d_{ab} \bar Q_{aL} 
\phi 
d_{bR} 
+ g^u_{ab}\bar Q_{aL} \tilde \phi u_{bR} + H.c.,
\label{11}
\end{equation}
where $a,b=1,2,3$.

In order to be this Lagrangian  $U(1)_Y $ gauge invariant we have:
\begin{eqnarray}
&&Y_{e_a} = Y_{l_a} - Y_\phi,\nonumber \\
&&Y_{q_a} =Y_q,\,\,\, Y_{u_a} = Y_u,\,\,\,Y_{d_a} = Y_d,\nonumber \\
\mbox{with}\nonumber \\
&&Y_u = Y_q + Y_\phi ,\,\,\, Y_d = Y_q - Y_\phi.
\label{12}
\end{eqnarray}
After that we have only  two nontrivial anomaly constraints: 
\begin{equation}
[SU(2)_L]^2 U(1)_Y \,\,\,\mbox{and}\,\,\, [U(1)_Y]^3_Y .
\label{13}
\end{equation}
So we are with five free parameters from the classical constraints, 
Eqs.(\ref{12}), and only 
two equations from the quantum constraints, Eqs.(\ref{13}). This prevents us 
from  
obtaining the QECH. This is again the effect of dequantization. The hidden 
symmetry 
here is $U(1)_L $, with $L$ 
being one of the quantum numbers:  $L = L_e - L_\mu,L_e - L_\tau,L_\mu - 
L_\tau$\cite{3}. 

To restore the QECH in the SM with three families we need to 
introduce either right-handed neutrinos with Majorana mass-term\cite{3,8}, or 
another Higgs doublet\cite{8} or some neutral fermions\cite{9}.

\section{The quantization in a chiral bilepton gauge model}

In a chiral bilepton gauge model presented in the introduction the electric 
charge operator can be defined in a general form as:
\begin{eqnarray}
Q=\frac{1}{2}(\lambda _3 -\sqrt{3}\lambda_8)+bN  , 
\label{14}
\end{eqnarray}
with $N$ being the operator generator of the group  $U(1)_N $; 
$\lambda_3$  and  $\lambda_8 $ being the two diagonal Gell-Mann matrices.

In order to break the symmetry spontaneously and to give mass to the 
fermions,
three triplets and one sextet of Higgs  are introduced\cite{10,11,12}: 
\begin{eqnarray}
\eta \sim (\mbox{{\bf 1,3}},N_\eta )\,\,\,,\,\,\,\rho \sim (\mbox{{\bf1,3}},
N_\rho )\,\,\,,\,\,\,\chi \sim (\mbox{{\bf 1,3}},N_\chi )\,\,\,,\,\,\,S\sim 
(\mbox{{\bf 1,6}},N_S),  
\label{15}
\end{eqnarray}
with the corresponding $U(1)_N$ charges as unknown parameters.

To generate mass correctly, those Higgs  must acquire the 
following vacuum 
expectation values\cite{10,11,12}: 
\begin{eqnarray}
&\langle&\eta \rangle_0 \sim \left( 
\begin{array}{c}
v_\eta \\ 
0 \\ 
0
\end{array}
\right) \,\,\,,\,\,\,\langle \rho \rangle_0 \sim\left( 
\begin{array}{c}
0 \\ 
v_\rho \\ 
0
\end{array}
\right)  \nonumber \\
&\langle&\chi \rangle_0 \sim \left( 
\begin{array}{c}
0 \\ 
0 \\ 
v_\chi
\end{array}
\right) ,\quad \langle S \rangle_0  \sim\left( 
\begin{array}{lcr}
0 & 0 & 0 \\ 
0 & 0 & v^{\prime } \\ 
0 & v^{\prime } & 0
\end{array}
\right) .  
\label{16}
\end{eqnarray}

With the requirement that the charge operator must annihilate the 
vacuum, we
obtain the following relations: 
\begin{equation}
 N_\eta =0,\,\,\,  b=\frac{1}{N_\rho },\,\,\,N_\chi =-N_\rho 
,\,\,\,N_S=0.  
\label{17}
\end{equation}

With these results we can write the electric charge operator in the 
following 
form: 
\begin{equation}
Q= I_3 +\frac{Y}{2},  
\label{18}
\end{equation}
with the hypercharge $Y$ being:
\begin{equation}
\frac{Y}{2}=I_8 +\frac{N}{N_\rho },
\label{19}
\end{equation}
where $I_3=\frac{1}{2}\lambda_3$  and  $I_8=-\frac{\sqrt{3}}{2}\lambda_8$.

Now the problem of the quantization of the electric charge consists in 
writing all the $U(1)_N$ charges as functions of $N_\rho$ . We 
achieve this 
in the same way as we did for the hypercharges in the SM, that is, by making 
use of 
the classical and quantum constraints. The 
only sector in the Lagrangian able to give information about the 
$U(1)_N$ charges through classical constraints is the Yukawa one, while the 
information about the $U(1)_N$ charges through quantum constraints comes from  
the three 
nontrivial 
anomaly cancellations: $[U(1_N)]^3$,  $[SU(3)_C]^2U(1)_N$  and  
$[SU(3)_L]^2U(1)_N$.

The leptons in the model come in the  $SU(3)$ representation:
\begin{equation}
L_{aL}=
\left (
\begin{array}{c}
\nu_a \\
e_a \\
e_a^c
\end{array}
\right )_L \,\, \sim (\mbox{{\bf 1,3}},N_{l_a})  ,
\label{20} 
\end{equation}
with $a=1,2,3$. The right-handed charged leptons enter the model through charge 
conjugation, i.e., 
there are no lepton singlets.

The quarks belong to $SU(3)$ and $U(1)$ representations and one family 
comes in 
triplets and the other ones in anti-triplets:
\begin{eqnarray}
&&Q_{1L}=
\left (
\begin{array}{c}
u_1 \\
d_1 \\
J_1
\end{array}
\right )_L \,\, \sim (\mbox{{\bf 3,3}},N_{q_1}),\nonumber \\
&&u_{1R} \sim (\mbox{{\bf 3, 1}},N_{u_1})\,\,\,\,\,\,d_{1R} \sim 
(\mbox{{\bf 
3,1}},N_{d_1})\,\,\,\,\,\, J_{R_1} \sim (\mbox{{\bf 3,1}},N_{J_1})  .
\label{21} 
\end{eqnarray} 
\begin{eqnarray}
&&Q_{iL}=
\left (
\begin{array}{c}
d_i \\
u_i \\
J_i
\end{array}
\right )_L \,\, \sim (\mbox{{\bf 3}}, \mbox{{\bf 
3}}^*,N_{q_i}),\nonumber \\
&&d_{iR} \sim (\mbox{{\bf 3,1}},N_{d_i}),\,\,\,\,\,\, u_{iR} \sim 
(\mbox{{\bf 
3,1}},N_{u_i}),\,\,\,\,\,\, J_{iR} \sim (\mbox{{\bf 3,1}},N_{J_i}) .
\label{22} 
\end{eqnarray}
Where $i=2,3$ .  The quarks $u$  e  $d$ are the usual ones with $J$ 
being 
the exotic quarks. Now we are ready to obtain the QECH.

In order to obtain relations among these $U(1)_N$ charges through the classical 
constraints, we use again the Yukawa Lagrangian sector\cite{12}: 
\begin{eqnarray}
-{\cal L}_{Y} &=&\frac{1}{2}G_{ab }\bar{L_{aL }^c}S^* L_{bL}  \nonumber \\
&+&\lambda _1\bar{Q}_{1L}J_{1R}\chi +\lambda _{ij}\bar{Q}_{iL}J_{jR}\chi 
^{*}
\nonumber \\
&+&\lambda^{\prime} _{1a}\bar{Q}_{1L}d_{aR}\rho +\lambda^{\prime} _{ia}\bar{Q}
_{iL}u_{aR}\rho^*  \nonumber \\
&+&\lambda^{\prime \prime} _{1a}\bar{Q}_{1L}u_{aR}\eta +\lambda^{\prime \prime} 
_{ia}\bar{Q}
_{iL}d_{aR}\eta ^{*}\,\,\,+\,\,\,H.c.,  \label{23}
\end{eqnarray}
where $a,\,\,b=1,\,2,\,3$ and $i,\,j=2,\,3$. $\chi ^{*}$ , $\rho ^{*}$ 
and 
$\eta^{*}$ are anti-triplets, while $S^{*}$ is anti-sextet.

The main point here is the leptonic sector of the Yukawa Lagrangian. Its 
framework provides the $U(1)_N$ charges of the leptons in a direct manner, 
that is, 
the $U(1)_N$ gauge invariance of this term implies: 
\begin{equation}
N_{l_1}=N_{l_2}=N_{l_3}=0.  
\label{24}
\end{equation}

The $U(1)_N$ gauge invariance of the quark sectors of the Yukawa 
Lagrangian 
implies: 
\begin{eqnarray}
&&N_{u_1}=N_{u_2}=N_{u_3}=N_u,\nonumber \\
&&N_{d_1}=N_{d_2}=N_{d_3}=N_d,  \nonumber \\
&&N_{q_2}=N_{q_3}=N_q,  \nonumber \\
&&N_{J_2}=N_{J_3}=N_J,\nonumber \\
\mbox{and},\nonumber \\
&&N_J=N_q-N_\rho ,  \nonumber \\
&&N_d=N_q,  \nonumber \\
&&N_u=N_q+N_\rho ,  \nonumber \\
&&N_{J_1}=N_q+2N_\rho ,  \nonumber \\
&&N_{q_1}=N_q+N_\rho .
\label{25}
\end{eqnarray}

After that we have  only one nontrivial anomaly 
cancellation\cite{13}: 
\begin{eqnarray}
[SU(3)_L]^2U(1)_N\Longrightarrow N_{q_1}+2N_q=0.  
\label{26}
\end{eqnarray}

From the last term in Eqs.(\ref{25}) and  (\ref{26}) we find: 
\begin{eqnarray}
N_q=-\frac{1}{3}N_\rho ,  
\label{27}
\end{eqnarray}
which leads  to  the following relations among the $U(1)_N$ charges: 
\begin{eqnarray}
&&N_{q_1}=\frac{2}{3}N_\rho ,\,\,\, N_q=-\frac{1}{3}N_\rho ,\,\,\,
N_{u}=\frac{2}{3}N_\rho ,\nonumber \\
&&N_{d}=-\frac{1}{3}N_\rho ,\,\,\, N_{J1}=\frac{5}{3}N_\rho ,\,\,\, 
N_{J}=-\frac{4}{3}N_\rho . 
 \label{28}
\end{eqnarray}

This result  allows us, together  with Eqs.(\ref{24}) and 
(\ref{19}) to 
find (when we replace
the value of the corresponding $U(1)_N$ charges) the hypercharges of 
all fermions. Substituting the hypercharges into Eq.(\ref{18}) we find the 
quantization of 
the electric charge with the correct electric charges for leptons and quarks: 
\begin{eqnarray}
&&Q_{\nu _e,\nu _\mu ,\nu _\tau }=0,  \nonumber \\
&&Q_{e,\mu ,\tau }=Q=-1,  \nonumber \\
&&Q_u=-\frac{2}{3},  \nonumber \\
&&Q_d=\frac{1}{3},  \nonumber \\
&&Q_{J1}=-\frac{5}{3},  \nonumber \\
&&Q_{J2,J3}=Q_J=\frac{4}{3}.  
\label{29}
\end{eqnarray}

Then, we have showed  that the chiral bilepton gauge model based on 
a semi-simple Lie group $G_{331}$ contains in its framework the quantization of 
the electric charge when we take into account the three families of fermions 
with massless neutrinos. 

We finish this section making a short analysis of the extensions of this model
in order to 
consider massive neutrinos. For having Majorana neutrinos we only need to 
use the following vacuum espectation value for the sextet\cite{12}:
\begin{eqnarray}
\langle S \rangle_0  \sim\left( 
\begin{array}{lcr}
v & 0 & 0 \\ 
0 & 0 & v^{\prime } \\ 
0 & v^{\prime } & 0
\end{array}
\right) .  
\label{30}
\end{eqnarray}
This conserves the structure of the leptonic sector, leading also to the 
quantization condition. Now if we want a Dirac neutrino we need to add the 
following term $G^{\prime}_{ab}\bar L_{aL} \eta \nu_{aR}$
to the Lagrangian ${\cal L}_Y$ in Eq.(\ref{23}) with $\nu_{aR} \sim 
(\mbox{{\bf 1, 1}},N_{\nu_a})$. By gauge invariance we find $N_{\nu_a}=0$. This 
result maintains the quantization condition. All this shows that this model is 
interesting in order to look for new physics.

\section{Conclusions}

Summarizing, we have extended the recent approach of the electric charge 
quantization problem in gauge models that contain an explicit $U(1)$ gauge group
to the case of one $G_{331}$ model. First we have
showed through classical and quantum constraints that the standard model with 
one family explains the quantization of the electric charge, while with three 
families and massless neutrinos it does not explain the charge quantization any 
more. We have  discussed the reasons for the above result and we have also 
showed that by adding neutrinos with
a Majorana mass to the 
standard model we restore the condition of the electric charge quantization.
 This happens because 
Majorana neutrinos break the $U(1)_L$ hidden symmetry which arises when we 
consider the 
standard model with three families. 

The central part of this work was to analyze the question of the quantization
of the electric charge in a $ G_{331}$ model and the main result in this work is 
the following: the QECH through classical and
quantum constraints occurs in the $G_{331}$ model when the three families are
taken together even if neutrinos are massless or not. If they are massive the
QECH does not depend on the nature of the neutrino fields i,e., it does not
matter if they are Dirac or Majorana fermions\cite{14}.

\acknowledgements
We thank J.C. Montero and V. Pleitez for the incentive and critical
suggestions and thank also to M. C. Tijero for reading the manuscript. This 
work was supported by the Conselho Nacional de
Desenvolvimento Cient\'{\i}fico e Tecnol\'{o}gico(CNPq)( O.P. R) and
Coordena\c {c}\~{a}o de Aperfei\c {c}oamento de Pessoal de N\'{\i}vel
Superior(CAPES)( C.A.S. P).


\begin{thebibliography}{99}
\bibitem{14} Unlike the standard model , the quantization of the electric
charge in the $G_{331}$ model is unavoidable.
\end{thebibliography}
\end{document}